\documentclass[notitlepage,aps,prl,reprint,superscriptaddress]{revtex4-1}

\renewcommand{\figurename}{FIG.}

\usepackage{mathtools}
\usepackage[english]{babel}
\usepackage[utf8]{inputenc}
\usepackage{amsmath}
\usepackage{amsfonts}
\usepackage{amssymb}
\usepackage{graphicx}
\usepackage{bbold}
\usepackage{epstopdf}
\usepackage{hyperref}
\renewcommand*{\Re}{\operatorname{Re}} 
\renewcommand*{\Im}{\operatorname{Im}} 
\DeclarePairedDelimiter\bra{\langle}{\rvert}
\DeclarePairedDelimiter\ket{\lvert}{\rangle}
\DeclarePairedDelimiter\bket{\big\lvert}{\big\rangle}

\begin{document}
\renewcommand{\figurename}{FIG.}
\renewcommand{\tablename}{TABLE}
\title{Measurement-free preparation of grid states}
\author{Jacob Hastrup}
\email{jhast@fysik.dtu.dk}
\affiliation{Center for Macroscopic Quantum States (bigQ), Department of Physics, Technical University of Denmark, Building 307, Fysikvej, 2800 Kgs. Lyngby, Denmark}
\author{Kimin Park}
\affiliation{Center for Macroscopic Quantum States (bigQ), Department of Physics, Technical University of Denmark, Building 307, Fysikvej, 2800 Kgs. Lyngby, Denmark}
\affiliation{Department of Optics, Palacky Univeristy, 77146 Olomouc, Czech Republic}
\author{Jonatan Bohr Brask}
\affiliation{Center for Macroscopic Quantum States (bigQ), Department of Physics, Technical University of Denmark, Building 307, Fysikvej, 2800 Kgs. Lyngby, Denmark}
\author{Radim Filip}
\affiliation{Department of Optics, Palacky Univeristy, 77146 Olomouc, Czech Republic}
\author{Ulrik Lund Andersen}
\affiliation{Center for Macroscopic Quantum States (bigQ), Department of Physics, Technical University of Denmark, Building 307, Fysikvej, 2800 Kgs. Lyngby, Denmark}

\begin{abstract}
Quantum computing potentially offers exponential speed-ups over classical computing for certain tasks. A central, outstanding challenge to making quantum computing practical is to achieve fault tolerance, meaning that computations of any length or size can be realised in the presence of noise. The Gottesman-Kitaev-Preskill code is a promising approach towards fault-tolerant quantum computing, encoding logical qubits into grid states of harmonic oscillators. However, for the code to be fault tolerant, the quality of the grid states has to be extremely high. Approximate grid states have recently been realized experimentally, but their quality is still insufficient for fault tolerance. Current implementable protocols for generating grid states rely on measurements of ancillary qubits combined with either postselection or feed forward. Implementing such measurements take up significant time during which the states decohere, thus limiting their quality. Here we propose a measurement-free preparation protocol which deterministically prepares arbitrary logical grid states with a rectangular or hexagonal lattice. The protocol can be readily implemented in trapped-ion or superconducting-circuit platforms to generate high-quality grid states using only a few interactions, even with the noise levels found in current systems.
\end{abstract}
\date{\today}

\maketitle

\section{INTRODUCTION}
Quantum computing offers exponential speeds-ups in solving certain computational problems, with wide-ranging consequences for information processing, information security, fundamental physics and chemistry and more. Impressive progress has been achieved towards realising quantum computing, including recent experimental demonstration of a quantum advantage over classical computation \cite{arute2019quantum}. However, real devices are subject to noise and imperfections. As computations grow in size and complexity, errors accumulate and eventually destroy any quantum advantage unless mitigated. Achieving fault tolerance, where errors are corrected sufficiently fast to allow scalable computation, is a central challenge to making universal quantum computing practical.

Quantum error correction (QEC) enables large-scale quantum computing in the presence of noise by redundantly encoding logical qubits into a larger Hilbert space. In traditional, discrete-variable QEC many physical qubits make up a single logical qubit. However, in 2001 Gottesman, Kitaev and Preskill (GKP) proposed encoding a logical qubit into the infinite-dimensional Hilbert space spanned by the continuous variables of a single bosonic mode \cite{gottesman2001encoding}. With this encoding, small displacement errors of the bosonic mode can be detected and corrected using only simple Gaussian operations. Furthermore, recent results have shown that the GKP code also performs very well against boson loss \cite{albert2018performance}, in  many cases outperforming other bosonic codes designed specifically against loss such as cat codes \cite{leghtas2013hardware,ofek2016extending} and binomial codes \cite{michael2016new,hu2019quantum}. In fact, numerical optimization suggests that the hexagonal GKP code might be the optimal loss-resistant code among all bosonic codes \cite{noh2018quantum}. Additionally, GKP codes have recently been shown to have applications within continuous-variable QEC \cite{noh2019encoding} and quantum metrology \cite{duivenvoorden2017single}.

An ideal GKP code is embedded in an idealized grid state which forms a lattice structure, consisting of an infinite superposition of position eigenstates. Such states require infinite energy and are hence unphysical. Importantly, however, it is possible to use approximate grid states with finite energy, composed of finitely squeezed states to achieve fault tolerance by concatenating the GKP code with discrete-variable error-correcting codes, provided that the grid states are sufficiently quadrature squeezed. In 2014, a conservative threshold for fault tolerance of 20.5 dB squeezing was derived for a  measurement-based quantum computing approach \cite{menicucci2014fault}. Later this threshold was significantly reduced to less than 10 dB squeezing by exploiting the analog information contained in the syndrome measurements \cite{fukui2018high,fukui2019high}. Other approaches such as concatenating the GKP code with the surface code \cite{noh2019fault}, the toric code \cite{wang2019quantum,vuillot2019quantum} and Knill's C$_4$/C$_6$ code \cite{fukui2017analog} have recently been proposed. For any of these proposals the squeezing threshold will depend not only on the involved codes, but also on the type and magnitude of the noise and experimental errors of the given system. It is therefore crucial to test the feasibility of these approaches with high-quality grid states experimentally. Additionally, as with any quantum error-correcting code, one would ideally use grid states with squeezing levels well above the threshold to avoid impractical resource overheads associated with the repeated concatenation of the codes. 

The preparation of grid states have, however, proven to be highly challenging. Recently, such states were prepared for the first time in ground-breaking experiments in the motional state of a trapped ion \cite{fluhmann2019encoding} and in a microwave cavity field coupled to a superconducting circuit \cite{campagne2019stabilized}. The states realised in these experiments clearly exhibit the required grid structure in phase space. However, the quality of the states needs to be improved for implementation with fault tolerant schemes. The main experimental limitation is that during the preparation protocol, the states accumulate noise e.g. from boson dephasing and losses, rendering the produced grid states noisy. To minimize this noise one has to increase the speed of the preparation protocol. The state-preparation protocols currently implemented in experiment use oscillator-qubit couplings and rely on repeated measurements of the ancilla qubit. These measurements and their associated processing times constitute about half of the total preparation time. Therefore, to improve the quality of the GKP codes, it is crucial to replace the slow measurement-based approach with a faster approach.  

In this work, we present a new, measurement-free grid-state preparation protocol, which is significantly faster than known methods, without introducing additional resources. The key interaction of our protocol is the Rabi interaction Hamiltonian between an oscillator and a two-level system \cite{kockum2019ultrastrong,forn2019ultrastrong}, which can be effectively simulated in trapped-ion and microwave systems. This interaction is also used in the experiments of Refs.~\cite{fluhmann2019encoding,campagne2019stabilized}. Such interactions were recently shown to enable deterministic, non-Gaussian operations by using many weak interactions \cite{park2017qubit,park2018deterministic}. Here, we instead use only a few, but stronger interactions, to generate the highly non-Gaussian grid states. Our work thus provides further demonstration of Rabi interactions as a powerful and versatile non-Gaussian resource in trapped ion and superconducting circuit platforms. 

The speed-up obtained with our approach is large enough to prepare grid states with more than $10$ dB of effective squeezing in practical systems that are readily available in both trapped-ion and microwave cavity platforms. With a further reduction of noise levels in future experiments, our protocol enables the generation of grid states with squeezing levels well above the fault-tolerance threshold levels, thus facilitating scalable quantum computing.

\section{GRID STATES}
In this section we review the basic structure of grid states and the figures of merit used in this article. For a more extensive review, see e.g. Ref. \cite{tzitrin2019towards}. 

Bosonic modes of harmonic oscillators are associated with the creation and annihilation operators $\hat{a}$ and $\hat{a}^\dagger$ and  the corresponding dimensionless quadrature operators $\hat{X}=\frac{1}{\sqrt{2}}(\hat{a} + \hat{a}^\dagger)$ and $\hat{P}=\frac{1}{\sqrt{2}i}(\hat{a} - \hat{a}^\dagger)$ satisfying $[\hat{X},\hat{P}]=i$. The 2-dimensional code space of the GKP-code is defined in the common +1 eigenspace of the stabilizer operators
\begin{equation}
    \hat{S}_z=\hat{D}(\alpha) \qquad \textrm{and} \qquad \hat{S}_x=\hat{D}(\beta).
\end{equation}
Here $\hat{D}(x) = e^{x\hat{a}^\dagger-x^*\hat{a}} = e^{i\sqrt{2}(-\textrm{Re}(x)\hat{P} + \textrm{Im}(x)\hat{X})}$ is the displacement operator with displacement amplitude $x$, satisfying the commutation relation 
\begin{equation}
    [\hat{D}(\alpha),\hat{D}(\beta)]=2i\sin(\Im(\alpha \beta^*))\hat{D}(\alpha+\beta).
\end{equation} 
By choosing $\Im(\alpha\beta^*)=2\pi$ we ensure that the stabilizers commute, which enables the existence of simultaneous eigenstates. Furthermore we can define logical operators 
\begin{equation}
    \hat{Z}_L=\hat{D}\left(\frac{\alpha}{2}\right), \ \hat{X}_L=\hat{D}\left(\frac{\beta}{2}\right), \ \textrm{and} \ \hat{Y}_L=\hat{D}\left(\frac{\alpha+\beta}{2}\right),
    \end{equation}
which commute with the stabilizers and anti-commute with each other. The logical GKP qubit states, $\ket{0}_\textrm{GKP}$ and $\ket{1}_\textrm{GKP}$, are then defined as the $\pm 1$ eigenstates of $\hat{Z}_L$. These satisfy the expected logic $\hat{X}_L\ket{0}_{\textrm{GKP}}=\ket{1}_{\textrm{GKP}}$ and $\hat{X}_L\ket{1}_{\textrm{GKP}}=\ket{0}_{\textrm{GKP}}$. 

The relative directions and magnitude of $\alpha$ and $\beta$ determine the lattice of the corresponding grid states. For example, rectangular grid states are generated by $\alpha=i2\pi/\beta^*$. Further choosing $\beta=\sqrt{2\pi}$ yields the square grid states for which the code space is symmetric with respect to $X$ and $P$. Alternatively, choosing $\alpha=i\sqrt{\frac{4}{\sqrt{3}}\pi}$ and $\beta = e^{-i\frac{\pi}{3}}\alpha$ yields the hexagonal grid states. In the following we will consider only the square grid, returning to the case of rectangular and hexagonals grid in Section \ref{sec:arbitrary}. The (unnormalizable) ideal square grid states can be written as:
\begin{subequations} 
\begin{align}
\ket{0}_{\textrm{GKP}} &= \sum_{s \in \mathbb{Z}}\hat{D}\left(s\sqrt{2\pi}\right)\ket{X=0} \\
\ket{1}_{\textrm{GKP}} &= \sum_{s \in \mathbb{Z}}\hat{D}\left(\left(s+\frac{1}{2}\right)\sqrt{2\pi}\right)\ket{X=0}
\end{align}\label{eq:GKPexact}\end{subequations}
where $\ket{X=0}$ denotes the eigenstate of $\hat{X}$ with eigenvalue $0$ and $\mathbb{Z}$ denotes the set of integers. The ideal grid states are thus infinite superpositions of equidistant position eigenstates and their Wigner functions are an infinite grid of 2-dimensional delta-functions (see Fig. \ref{fig:result}(b)). Ideal grid states can be approximated by finite-energy states in several ways. The most commonly used representation for deriving fault tolerance thresholds is a superposition of finitely squeezed states of width $e^{-r}$ under a Gaussian envelope of width $\kappa^{-1}$:
\begin{subequations}
\begin{align}
&\ket{\tilde{0}}_{\textrm{GKP}} \propto \sum_{s \in \mathbb{Z}}e^{-\frac{\left(2\sqrt{\pi} s\right)^2}{2\kappa^{-2}}}\hat{D}\left(s\sqrt{2\pi}\right)\hat{S}_r\ket{\textrm{vac}} \\
&\ket{\tilde{1}}_{\textrm{GKP}} \propto \sum_{s \in \mathbb{Z}}e^{-\frac{\left(2\sqrt{\pi} \left(s+\frac{1}{2}\right)\right)^2}{2\kappa^{-2}}}\hat{D}\left(\left(s+\frac{1}{2}\right)\sqrt{2\pi}\right)\hat{S}_r\ket{\textrm{vac}}, 
\end{align}
\label{eq:GKPapprox}
\end{subequations}
where $\hat{S}_r=e^{-\frac{1}{2}r(\hat{a}^2-\hat{a}^{\dagger2})}$ is the squeezing operator (not to be confused with the stabilizers $\hat{S}_x$ and $\hat{S}_z$). The squeezing parameter $r$ and envelope $\kappa$ characterises the quality of the states in the $X$- and $P$-quadratures respectively and in the limit $(e^{-r},\kappa) \rightarrow (0,0)$ the approximate states converge to the exact states of equation \eqref{eq:GKPexact}. For $\kappa=e^{-r}$ the states can correct noise equally well in $X$ and $P$.

However, physical grid states will never exactly be of the form given in equation \eqref{eq:GKPapprox}. First, physical states are not pure and are generally described by a density matrix $\hat{\rho}$. Secondly, the exact Gaussian envelope can be difficult to obtain and most preparation protocols yield a finite sum of squeezed states. Therefore, the parameters $r$ and $\kappa$ are not well-defined for practically realizable states. 
Instead, more generic figures of merit, the \textit{effective squeezing parameters}, have been suggested in Ref. \cite{duivenvoorden2017single}. They quantify the effective degree of squeezing in each quadrature of the peaks constituting the grid state and are defined as:
\begin{subequations}
\begin{align}
\Delta_X &= \sqrt{\frac{1}{2\pi}\ln\left(\frac{1}{|\langle\hat{D}(i\sqrt{2\pi})\rangle|^2}\right)}\\
\Delta_P& = \sqrt{\frac{1}{2\pi}\ln\left(\frac{1}{|\langle\hat{D}(\sqrt{2\pi})\rangle|^2}\right)},
\end{align}
\label{eq:Delta}\end{subequations}
where the effective squeezing levels in units of dB are given by $\Delta_\textrm{dB}=-10\log_{10}(\Delta^2)$. The expectation values in these definitions are exactly the expectation values of the stabilizers $\hat{S}_z$ and $\hat{S}_x$ for square GKP states. High quality grid states should therefore have $|\langle\hat{S}_{z/x}\rangle|\approx 1$, in which case $\Delta_{X/P}\rightarrow\infty$ dB. These definitions also have the nice property that for squeezed states they reproduce the squeezing parameter, i.e. $\Delta_X(\hat{S}_r \ket{\textrm{vac}})=e^{-r}$ and $\Delta_P(\hat{S}_r\ket{\textrm{vac}})=e^r$. Furthermore, for approximate square lattice grid states of Eq. \eqref{eq:GKPapprox} we extract the parameters $\Delta_X(\ket{\tilde{0}}_\textrm{GKP})=e^{-r}$ and $\Delta_P(\ket{\tilde{0}}_\textrm{GKP})\approx\kappa$. The last approximation is very accurate for $e^{-r},\kappa>10$ dB. Thus, if the state resembles a grid state consisting of a grid of squeezed peaks in phase-space, the effective squeezing parameters approximately quantify the squeezing of these peaks in each quadrature direction. 

However, it is important to note that the effective squeezing is not directly related to the fault-tolerance thresholds. Most GKP-based fault-tolerance thresholds are derived based on the specific, approximate states given in Eq. \eqref{eq:GKPapprox} and refer to $r$ and $\kappa$. Any other state can therefore in general not be guaranteed to enable fault tolerant computations, even when the effective squeezing parameters are both above these thresholds. Moreover, the effective squeezing parameters say nothing about the logic state of the GKP qubit, e.g. a mixed code state might be strongly squeezed, but might not necessarily be useful for quantum computing. 

Nevertheless, it is reasonable to assume that states with a high degree of effective squeezing can be used for fault tolerance if they otherwise closely resemble the approximate pure grid states of Eq. \eqref{eq:GKPapprox}, e.g. in terms of their fidelity with the approximate states. In the further analysis we therefore compliment the effective squeezing parameters with the fidelity to verify the appropriateness of using the effective squeezing parameters as quantifiers of the protocol performance. Moreover, we also verify that the produced states have the expected grid structure in terms of their Wigner function. Another alternative figure of merit is the \textit{effective shift error} \cite{glancy2006error} which is discussed and calculated in Appendix \ref{app:shifterror}.
\begin{figure*}
\centering
\includegraphics{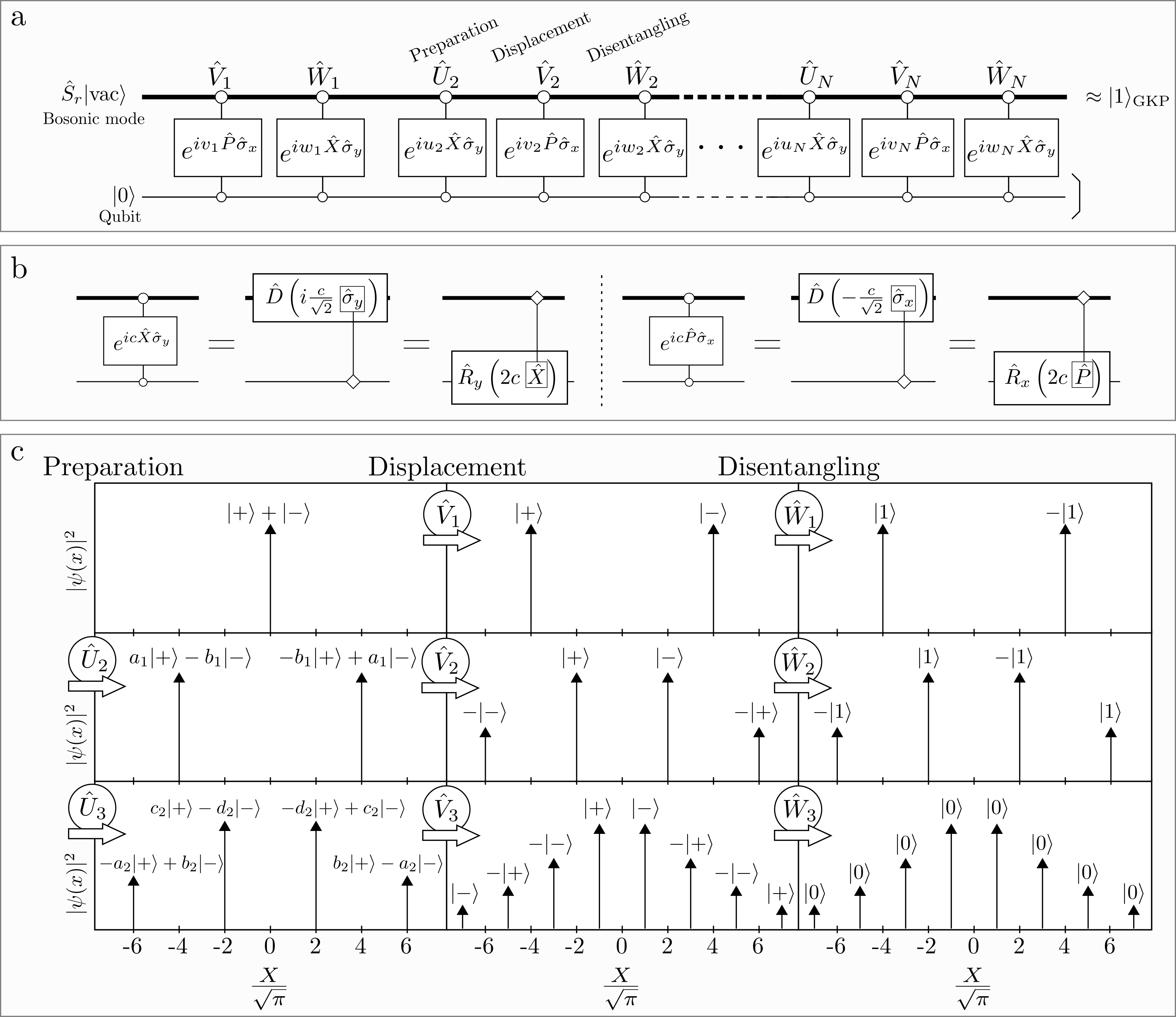}
\caption{(a) Circuit diagram of the measurement-free grid state generation protocol. The bosonic mode interacts with the qubit through a sequence of Rabi gates with interaction Hamiltonians of the form $\hat{P}\hat{\sigma}_x$ and $\hat{X}\hat{\sigma}_y$ to produce an approximate GKP 1 logic state without measurements. (b) The Rabi gates can be viewed either as conditional displacements on the bosonic mode depending on the qubit state or conditional rotations of the qubit depending on the bosonic state. (c) Illustration of the protocol for $N=3$ for an infinitely squeezed input state. The ket above each peak in the wave function represents the state of the qubit entangled with the given peak. The displacement gates $\hat{V}$ split each peak in two, creating an entangled state. The disentangling gates $\hat{W}$ then rotate the qubit depending on the boson state to remove the entanglement. The preparation gates $\hat{U}$ rotate the qubit before the displacement gates to control the envelope of the resulting state.}
\label{fig:scheme}
\end{figure*}

\section{PREPARATION PROTOCOL}
Several proposals exist for the preparation of approximate grid states \cite{gottesman2001encoding,travaglione2002preparing,campagne2019stabilized,terhal2016encoding,shi2019fault,su2019generation,eaton2019gottesman,vasconcelos2010all,weigand2018generating}. The original GKP paper \cite{gottesman2001encoding} includes a proposal based on an radiation-pressure-like interaction between two bosonic modes under the Hamiltonian $\hat{X}_1 \hat{a}_2^\dagger\hat{a}_2$ in the quantum non-linear regime. However, experimental realization of the required strongly nonlinear coupling has proven highly challenging and has not yet been achieved. 

In \cite{travaglione2002preparing}, a preparation protocol based on the Rabi interaction Hamiltonian $\hat{P} \hat{\sigma}_x$ (where $ \hat{\sigma}_x$
 is the Pauli-$x$ matrix), between the bosonic mode and a two-level system was proposed. Such an interaction can be realized in trapped ions \cite{haljan2005spin} and microwave cavities \cite{campagne2019stabilized}. This protocol, however, has three main drawbacks: First it is probabilistic, with a success probability inversely proportional to the mean photon number of the generated state. Secondly, the output states have a box-shaped envelope rather than the Gaussian envelope of equation \eqref{eq:GKPapprox}. This means that the effective squeezing parameters are suboptimal given the number steps required to prepare the states. Hence, excessively large states need to be generated to obtain useful effective squeezing. Finally, the protocol requires qubit measurements, which in realistic systems will constitute a significant contribution to the total preparation time during which the state decoheres. 
 
The two first issues were solved by Terhal and Weigand in Ref. \cite{terhal2016encoding}: By adding a single measurement-based feed-forward displacement operation as well as suitable qubit rotations, the protocol is made deterministic. Furthermore, by using a different strength of the Rabi interactions, the envelope of the output state is made nearly Gaussian, making the protocol much more efficient. However, their protocol still relies on qubit measurements, which limits the quality of the states that can be realistically generated in the laboratory today. 

Our protocol addresses all the above mentioned problems by adding additional short Rabi interactions of the form $\hat{X}\hat{\sigma}_y$, which effectively act as ``deterministic measurements" by disentangling the bosonic mode and the qubit and further enables us to shape the envelope of the state. This interaction can be obtained from the $\hat{P}\hat{\sigma}_x$ Hamiltonian by simple rotations of the qubit and the bosonic mode, Similarly, both interaction types can be obtained from the more commonly considered Rabi Hamiltonian $\hat{X}\hat{\sigma}_x$. Fig. \ref{fig:scheme}(a) shows a circuit diagram of the protocol. It consist of $N$ groups of interactions, each consisting of 3 gates:
\begin{itemize}
\item a \textit{preparation gate}, $\hat{U}_k=e^{iu_k\hat{X}\hat{\sigma}_y}$.
\item a \textit{displacement gate}, $\hat{V}_k=e^{iv_k\hat{P}\hat{\sigma}_x}$.
\item a \textit{disentangling gate}, $\hat{W}_k=e^{iw_k\hat{X}\hat{\sigma}_y}$.
\end{itemize}
These interactions can be interpreted as either conditional displacements of the bosonic mode or conditional rotations of the qubit, as illustrated in Fig. \ref{fig:scheme}(b). Since the preparation and disentangling gates are of the same type, i.e. $\hat{X}\hat{\sigma}_y$, the preparation gate of round $k$ can be combined with the disentangling gate of round $k-1$ into a single gate. The interaction strengths of the displacement and disentangling gates are given by
\begin{align}
v_k &=\begin{cases}
                -\sqrt{\pi}2^{N-1}, & \text{if $k=1$,}\\
               \sqrt{\pi}2^{N-k} & \text{if $k>1$.}\\
            \end{cases} \\
            w_k &=\begin{cases}
               -\frac{\sqrt{\pi}}{4 }2^{-(N-k)}, & \text{if $k<N$,}\\
               \frac{\sqrt{\pi}}{4 } & \text{if $k=N$.}\\
            \end{cases}
\end{align}
while the interaction strengths of the preparation gates, $u_k$, are found numerically (see Appendix \ref{app:prep}). In the first round the optimal preparation gate strength is $u_1=0$ i.e. $\hat{U}_1=\mathbb{1}$ so $\hat{U}_1$ is thus ignored in Fig. \ref{fig:scheme}(a). The input state is a squeezed vacuum state $\hat{S}_r\ket{\textrm{vac}}$ and the output state is an approximation to the state $\ket{1}_\textrm{GKP}$, which can subsequently be transformed into an arbitrary grid state, as will be discussed later. Note that all gates commute with $\hat{D}(i\sqrt{2\pi})=e^{i2\sqrt{\pi}\hat{X}}$. Therefore, $\Delta_X$ is left invariant under the protocol, i.e. the effective squeezing of the output state in the $X$-quadrature is $\Delta_X=e^{-r}$. The effect of the protocol is thus to create a superposition of $2^N$ squeezed states and thereby improve $\Delta_P$. The effect of each gate is illustrated in Fig. \ref{fig:scheme}(c) for the case of $N=3$, but the procedure can be extended for arbitrary $N$.

\subsection{Infinitely squeezed input states}
To illustrate the functionality of the gates, we first consider an infinitely squeezed input state, $\ket{X=0}$. For brevity we will use the notation $\ket{X=x_0}=\ket{x_0}_x$ in the following. The first operation is the displacement gate $\hat{V}_1$ which creates an entangled boson-qubit state:
\begin{equation}
\hat{V_1}\ket{0}_x\ket{0}=\frac{1}{\sqrt{2}}\left(\bket{2^{N-1}\sqrt{\pi}}_x\ket{+} + \bket{-2^{N-1}\sqrt{\pi}}_x\ket{-} \right),
\end{equation}
where $\ket{\pm}=(\ket{0} \pm \ket{1})/\sqrt{2}$. The disentangling gate then rotates the qubit to erase the entanglement:
\begin{equation}
\hat{W_1}\hat{V_1}\ket{0}_x\ket{0}=\frac{1}{\sqrt{2}}\left(\bket{2^{N-1}\sqrt{\pi}}_x - \bket{-2^{N-1}\sqrt{\pi}}_x \right)\ket{1}.
\end{equation}
We have thus created a superposition between two squeezed states. The second round splits each of these peaks in two, creating a total of four peaks:
\begin{align}
&(\hat{W_2}\hat{V_2}\hat{U_2})(\hat{W_1}\hat{V_1})\ket{0}_x\ket{0}  \nonumber \\ 
=&\frac{1}{\sqrt{2}}\big(-b_1\bket{-3\cdot2^{N-2}\sqrt{\pi}}_x + a_1\bket{-2^{N-2}\sqrt{\pi}}_x \nonumber\\
&-a_1\bket{2^{N-2}\sqrt{\pi}}_x + b_1\bket{3\cdot2^{N-2}\sqrt{\pi}}_x\big)\ket{1}.
\end{align}
The coefficients are controlled by the preparation gate and are given by $a_1=\sin(\pi/4 + 2^{N-1}\sqrt{\pi}u_2)$ and $b_1=\cos(\pi/4 + 2^{N-1}\sqrt{\pi}u_2)$. The third round creates 8 peaks and so on for a total of $2^N$ peaks after $N$ rounds. Thus, the resulting state is
\begin{equation}
\left(\frac{1}{\sqrt{2}}\sum_{k=1}^{2^N}c_k\bket{\left(2k-2^N-1\right)\sqrt{\pi}}_x\right)\ket{0}, \label{eq:result}
\end{equation}
where the coefficients $c_k$ can be optimized by tuning the strengths of the preparation gates (see Appendix \ref{app:prep}). For these infinitely squeezed input states we can obtain $\Delta_P=\{6.6,11.6,16.6,20.6\}$ dB, for $N=\{1,2,3,4\}$, and $\Delta_X=\infty$ dB as $\Delta_X$ is determined solely by the initial squeezing of the input state. 

\begin{figure}
\includegraphics{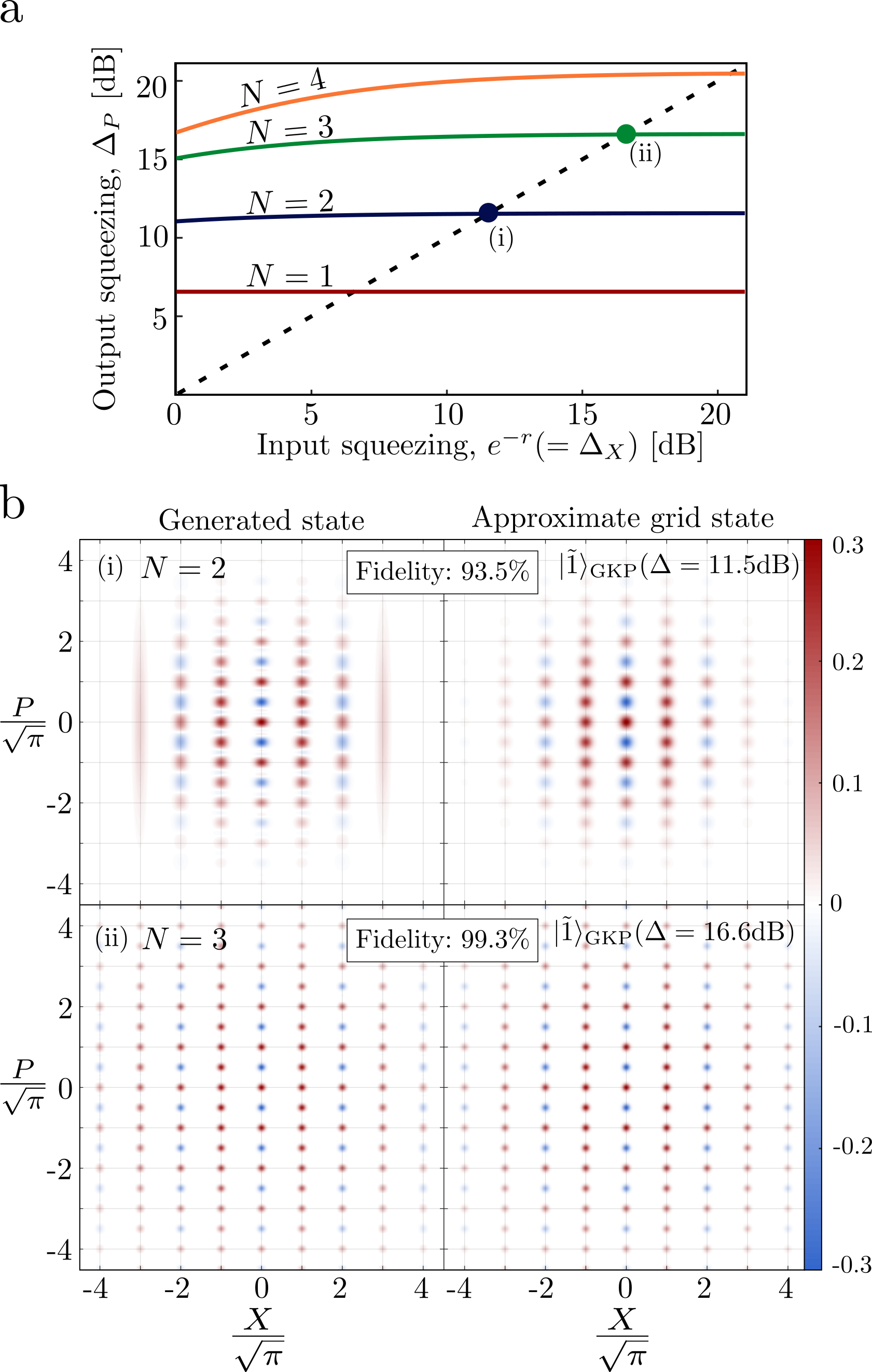}
\caption{(a) Effective squeezing in the $P$ quadrature, $\Delta_P$, as a function of squeezing in the $X$ quadrature of the input state after $N$ rounds. $\Delta_X$ is invariant under the protocol. The dashed line is $\Delta_P=\Delta_X=e^{-r}$. (b) Left: Wigner functions of the generated states for $N=2$ and $N=3$ with $11.5$ dB and $16.6$ dB input squeezing  respectively. The plotted states are marked with (i) and (ii) in (a). Right: Wigner functions of the target approximate GKP states given by equation \eqref{eq:GKPapprox}.}
\label{fig:result}
\end{figure}

\subsection{Finitely squeezed input states}
For a finitely squeezed input state the protocol outlined above is not exact, and in particular the disentangling operation is not exact. Thus, after tracing out the qubit the resulting state is mixed, but the effect on $\Delta_P$ of the output state is very small. This can be seen in Fig. \ref{fig:result}(a) which shows $\Delta_P$ as a function of the input squeezing. Since $\Delta_X=e^{-r}$ is preserved during the protocol, high effective squeezing can be obtained simultaneously in both quadratures even with finitely squeezed input states. 

Note that, as seen from Fig. \ref{fig:result}, even with vacuum input a significant amount of effective squeezing can be obtained. By applying the protocol twice, once in each quadrature direction, we can therefore generate grid-like states with high degrees of effective squeezing in both quadratures. However, a careful analysis (presented in Appendix \ref{app:nosqueeze}) shows that these states are not well-defined pure states in the GKP basis, and therefore, seemingly, unsuitable for GKP-based computations.

In Fig. \ref{fig:result}(b) (left) we present the Wigner functions of the generated states for $N=2$ and $N=3$ with input squeezing of $11.5$ dB and $16.6$ dB respectively, in which case equal effective squeezing in $X$ and $P$ is obtained. For comparison, we also plot the Wigner functions of the corresponding target approximate grid states given by equation \eqref{eq:GKPapprox} with the same amount of squeezing (right plots in Fig. \ref{fig:result}(b)). For $N=2$ we observe very small differences in the edges of the states which are caused by the cut-off in the number of squeezed states in the superposition of the generated state. Despite these differences, the resulting fidelity is already $93.5\%$. For $N=3$ the differences becomes much less pronounced and the fidelity increases to $99.3\%$. Thus, very few rounds of operations are required to make grid states with high effective squeezing and near unity fidelity to the commonly considered approximate grid states of equation \eqref{eq:GKPapprox}.

\begin{figure}
\includegraphics{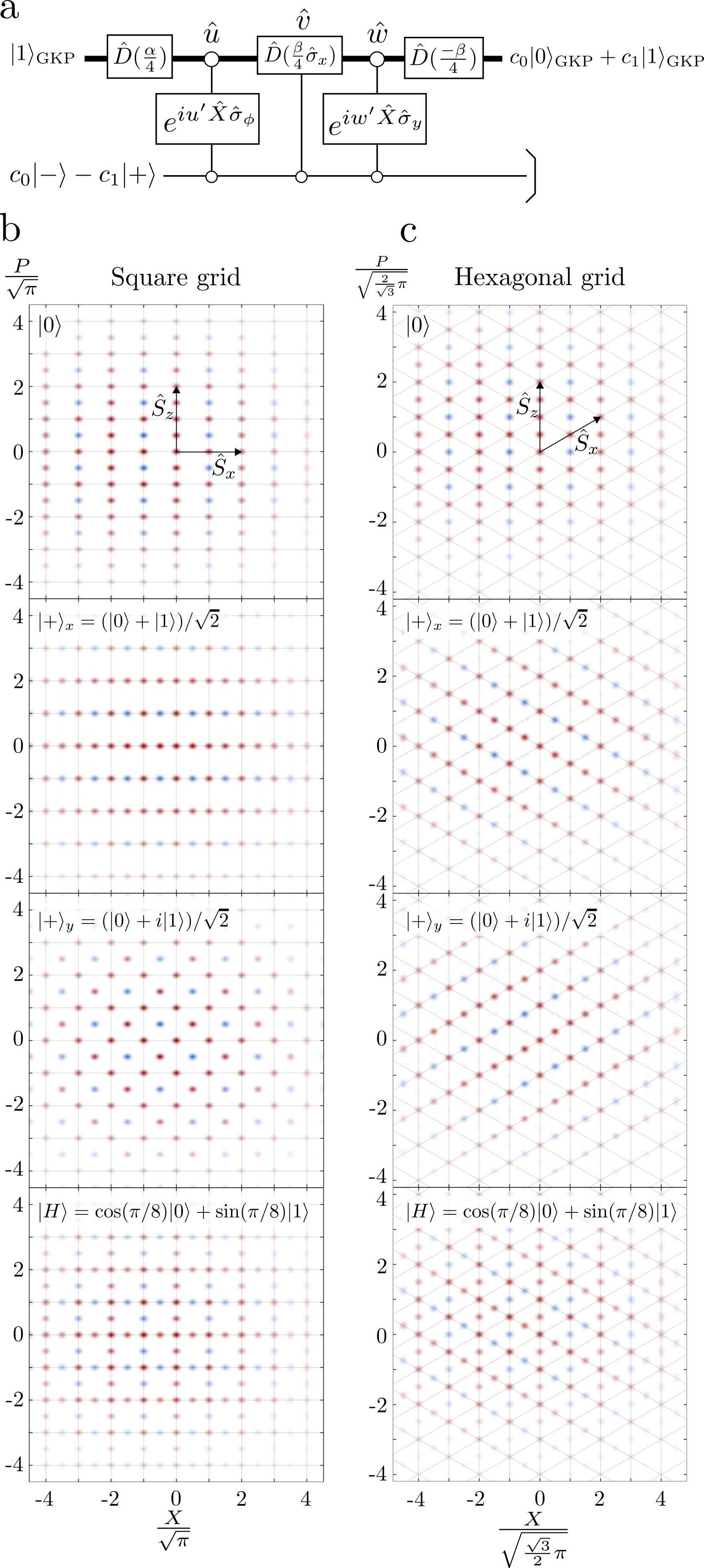}
\caption{(a) Circuit diagram for preparing arbitrary logical GKP states. (b) Wigner functions for various logical square grid states numerically generated using our scheme with $N=3$ and 16.6 dB input squeezing. (c) Wigner functions for various logical hexagonal grid states generated using $N=3$ and 15 dB input squeezing. The arrows in the top plot of (b) and (c) show the directions and magnitude of the stabilizer displacements $\hat{S}_z$ and $\hat{S}_x$.}
\label{fig:logicalStates}
\end{figure}

\subsection{Preparation of arbitrary logical states}\label{sec:arbitrary}
The state generated so far is the logical $\ket{1}$ state of the square GKP code. It is, however, important to be able to generate an arbitrary logical grid state, i.e. of the form $c_0\ket{0}_\textrm{GKP} + c_1\ket{1}_\textrm{GKP}$. In particular, magic states, such as $\ket{H}=\cos(\pi/8)\ket{0}_\textrm{GKP} + \sin(\pi/8)\ket{1}_\textrm{GKP}$ are highly important as they serve as resources for performing non-Clifford operations via gate teleportation.

Furthermore, non-square rectangular grid states--which are equivalent to squeezed square grid states--are also a useful resource, as they remove the need for in-line squeezing using a newly developed modified Glancy and Knill error recovery scheme \cite{wan2019memory}. In the following we thus discuss how to generate the arbitrary logical grid state with both rectangular and hexagonal lattices.

We first note that rectangular lattices map onto square lattices simply by scaling the quadratures, i.e. $\hat{X}\rightarrow C \hat{X}$ and $\hat{P}\rightarrow C^{-1} \hat{P}$ where $C$ is the scale factor. These scalings can consequently be straightforwardly implemented by appropriate scaling of the interaction parameters, i.e. $u\rightarrow C u$, $v\rightarrow C^{-1} v$ and $w\rightarrow C w$. To generate hexagonal states we utilize the fact that the hexagonal logical $\ket{1}$ state is identical to the logical $\ket{1}$ state of the rectangular lattice with $\alpha = i\sqrt{\frac{4}{\sqrt{3}}\pi}$ and $\beta = \sqrt{\sqrt{3}\pi}$. We can thus also initialize the logical $\ket{1}$ state of the hexagonal lattice.

The circuit diagram shown in Fig. \ref{fig:logicalStates}(a) shows how to map the logical $\ket{1}$ state into arbitrary logic states using three Rabi-interactions $\hat{u}$, $\hat{v}$ and $\hat{w}$. The idea is to proceed with the scheme for generating the logical $\ket{1}$ state, but exploiting the linearity of the protocol and the fact that the effect of the displacement gate depends on the state of the qubit. Therefore, by initializing the qubit in the state $c_0\ket{-} - c_1\ket{+}$, we effectively transfer the coefficients of the qubit onto the grid state. The additional two, unconditional, displacement operations ensure that the resulting state is on the lattice. The first operation $\hat{D}(\alpha/4)$ can be effectively implemented during the preparation protocol by inverting the sign of $w_N$, while the second unconditional operation $\hat{D}(\beta/4)$ simply shifts the lattice for all states and can therefore be virtually implemented by a shift of reference frame. The strength of the first conditional operation $u'$, and the qubit-dependence $\phi$ of the preparation gate $\hat{u}$ depends on the target logical state and are found by numerical optimization ($\hat{\sigma}_\phi=\cos(\phi)\hat{\sigma}_x + \sin(\phi)\hat{\sigma}_y$ represents a generalized Pauli operator in the $x$--$z$ plane). The gate $\hat{u}$ is not crucial for the scheme, but only improves the quality of the output states by allowing a degree of control over the envelope of the output state. The strength of the disentangling gate $\hat{w}$ is $w' = -\pi/(\sqrt{2}\textrm{Re}(\beta))$. 

Fig. 3(b) and (c) show the Wigner functions of various logical states with square and hexagonal lattices respectively, numerically generated using this protocol, showing clear, well-defined grid structures. 

\begin{figure}
\includegraphics{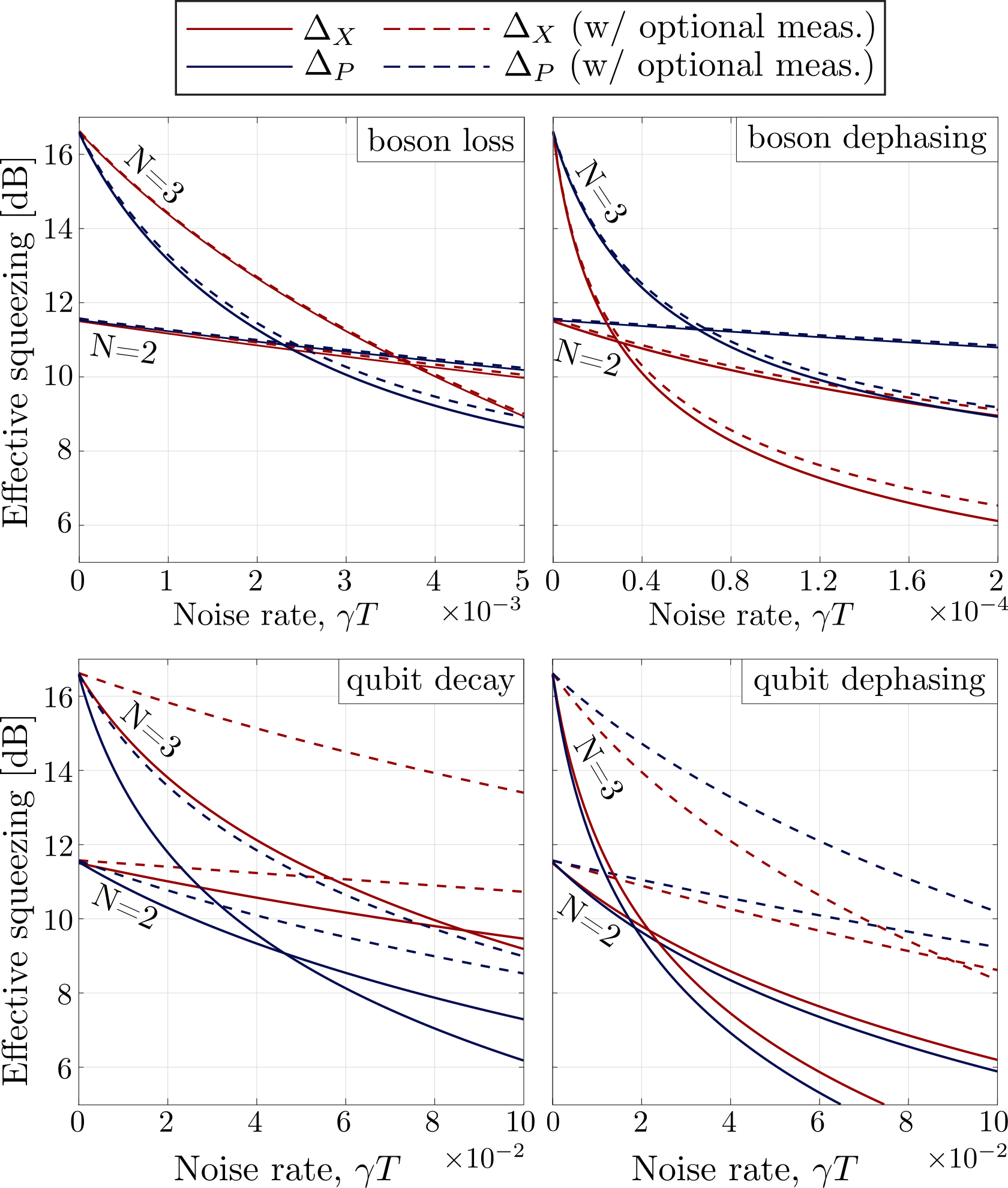}
\caption{Effective squeezing in $X$ and $P$ for a square grid state as a function of noise rate for different noise sources during the preparation protocol. $\gamma$ is the noise rate and $T$ is the time required to implement $e^{i\hat{X}\hat{\sigma}_y}$ and $e^{i\hat{P}\hat{\sigma}_x}$.}
\label{fig:noise}
\end{figure}

\section{EFFECTS OF NOISE}
We now consider the effect of relevant noise sources on our protocol. To include noise effects in our model, we consider each gate as being implemented with a specific Hamiltonian for a set duration, e.g. the gate $e^{ic\hat{X}\hat{\sigma}_y}$ is implemented via the Hamiltonian $\hat{H}=\frac{1}{T}\hat{X}\hat{\sigma}_y$ within the time $t = cT$. To simulate the added noise, we use a master equation approach in which noise is included in the Lindblad terms $\hat{L}$:
\begin{equation}
    \frac{\textrm{d}\rho}{\textrm{d}t}=-\frac{i}{\hbar}[\hat{H},\rho] + \hat{L}\rho\hat{L}^\dagger - \frac{1}{2}\left(\hat{L}^\dagger\hat{L}\rho+\rho\hat{L}^\dagger\hat{L}\right),
    \label{eq:master}
\end{equation}
where $\rho$ is the density matrix of the composite boson-qubit system. We consider four common noise channels:
\begin{itemize}
\item Boson loss: $\hat{L}=\sqrt{\gamma}\hat{a}$
\item Boson dephasing: $\hat{L}=\sqrt{\gamma}(\hat{a}\hat{a}^\dagger + \hat{a}^\dagger\hat{a})$
\item Qubit dephasing: $\hat{L}=\sqrt{\gamma}\hat{\sigma}_z$\\
\item Qubit decay: $\hat{L}=\sqrt{\gamma}(\hat{\sigma}_x + i\hat{\sigma}_y)/2$
\end{itemize} 
The effect of these noise sources on the effective squeezing of the output states is shown by the solid lines in Fig. \ref{fig:noise}. For each noise source we consider $N=2$ and $N=3$ rounds with $11.5$ dB and $16.6$ dB squeezed input states respectively. It is clear that our protocol is sensitive to all types of noise. By increasing $N$ we also increase the implementation time of the protocol, thus increasing the effect of noise. Therefore, there exists an optimal number of rounds that depends on the magnitude and type of noise. E.g. for large noise contributions, two rounds ($N=2$) of the scheme produces states with higher effective squeezing degrees than three rounds ($N=3$), and this is simply a result of the extended time over which noise can accumulate. This clearly illustrates the importance of a fast preparation protocol. 

Even though the quality of the generated states is limited by qubit and bosonic errors, the effect of qubit errors can be significantly suppressed by adding a few qubit measurements, after each of the disentangling gates $\hat{W}$. In the noiseless case, the qubit should be in a known state, disentangled from the bosonic mode at these points, as illustrated in the rightmost windows of Fig. \ref{fig:scheme}(c). Therefore, if we measure the qubit in a different state, we know that an error has occurred, and the realization should be discarded and the protocol restarted. The result of such a postselection strategy is shown by the dashed lines in Fig. \ref{fig:noise}, demonstrating that we can improve the effective squeezing of the output state by several dB. Bosonic errors, on the other hand, are largely unaffected by the postselection strategy. Thus, when these errors are dominating the only way to improve the output states is to increase the interaction speed or reduce the rate of the noise. For the calculation of Fig. \ref{fig:noise} we assumed instantaneous measurements to isolate the effect of qubit projections. In real systems the measurements will take time, during which noise accumulates thus resulting in lower effective squeezing parameters. However, compared to the measurement-based schemes, e.g. phase-estimation \cite{terhal2016encoding}, we require exponentially fewer measurements and therefore still attain a significant speed-up.

Using realistic noise parameters and operation speeds from recent experiments with trapped ions \cite{fluhmann2019encoding} and microwave cavities \cite{campagne2019stabilized}, we find that grid states with effective squeezing parameters above 10 dB in both quadrature can be realistically generated in both platforms using input states squeezed by 11 dB (see Appendix \ref{app:realisticnoise}). Squeezing levels of 12.6 dB in trapped ions \cite{kienzler2015quantum} and 10 dB in microwave cavities \cite{castellanos2008amplification} have been experimentally generated. 

\section{Conclusion}
In conclusion, we have presented a measurement-free protocol to deterministically prepare GKP states using only few interactions of the type $\hat{X}\hat{\sigma}_{y}$ and $\hat{P}\hat{\sigma}_x$, which are readily available in trapped-ion and microwave-cavity platforms. Our protocol requires no measurements, resulting in a speed-up over previous methods, which enables the generation of grid states with high effective squeezing levels. Furthermore, by adding a few measurements we can partly detect qubit errors, thus making the protocol robust against qubit noise. Although the exact requirements for general CV states (i.e. states not exactly on the form of Eq. \eqref{eq:GKPapprox}) to enable fault-tolerance with the GKP encoding are yet unknown, it seems reasonable that states generated using this protocol suffices, due to their high fidelity with the commonly considered approximate grid states of Eq. \eqref{eq:GKPapprox}. 

Finally, our protocol exemplifies the versatility of sequential applications of non-commuting Rabi Hamiltonians, e.g. $\hat{P}\hat{\sigma}_x$ and $\hat{X}\hat{\sigma}_y$, demonstrating that highly non-Gaussian states can be deterministically engineered with only a few of these interactions. The full power of such repeated combination of Rabi interactions remains still relatively unexplored, but we expect that many other interesting applications are possible using this technique.

\section*{Acknowledgements}
\begin{acknowledgements}
This project was supported by the Danish National Research Foundation through the Center of Excellence for Macroscopic Quantum States (bigQ). R.F. acknowledges project LTAUSA19099 and 8C20002 from the Ministry of Education, Youth and Sports of Czech Republic. K.P. acknowledges Project GB19-19722J of the Czech Science Foundation.
\end{acknowledgements}

\newpage

\renewcommand{\appendixname}{APPENDIX}
\begin{appendix}
\section{EFFECTIVE SHIFT ERROR}\label{app:shifterror}
In \cite{glancy2006error} it was show that small displacement errors on ideal grid states can be perfectly corrected, if the magnitude of the displacements are less than $\sqrt{\pi}/6$. Furthermore, any bosonic state $\rho$ can be expanded in a basis of shifted ideal grid states:
\begin{align}
    \rho=&\int_{-\sqrt{\pi}}^{\sqrt{\pi}}du\int_{-\sqrt{\pi}}^{\sqrt{\pi}}du' \nonumber \\  &\times\int_{-\sqrt{\pi}/2}^{\sqrt{\pi}/2}dv\int_{-\sqrt{\pi}/2}^{\sqrt{\pi}/2}dv'\rho_{uv,u'v'}\ket{u,v}\bra{u',v'},
    \end{align}
where
\begin{equation}
    \ket{u,v}=\pi^{-\frac{1}{4}}\hat{D}(u/\sqrt{2})\hat{D}(-iv/\sqrt{2})\ket{0}_\textrm{GKP}.
\end{equation}
The \textit{effective shift error}, defined as the probability that an approximate logical 0 GKP state has an intrinsic displacement error larger than $\sqrt{\pi}/6$ is then given by:
\begin{equation}
    P_{\textrm{error}}^{\sqrt{\pi}/6} = 1-\int_{-\sqrt{\pi}/6}^{\sqrt{\pi}/6}du\int_{-\sqrt{\pi}/6}^{\sqrt{\pi}/6}dv \rho_{uv,uv}.
    \label{eq:Perror}
\end{equation}

\begin{figure}
    \centering
    \includegraphics{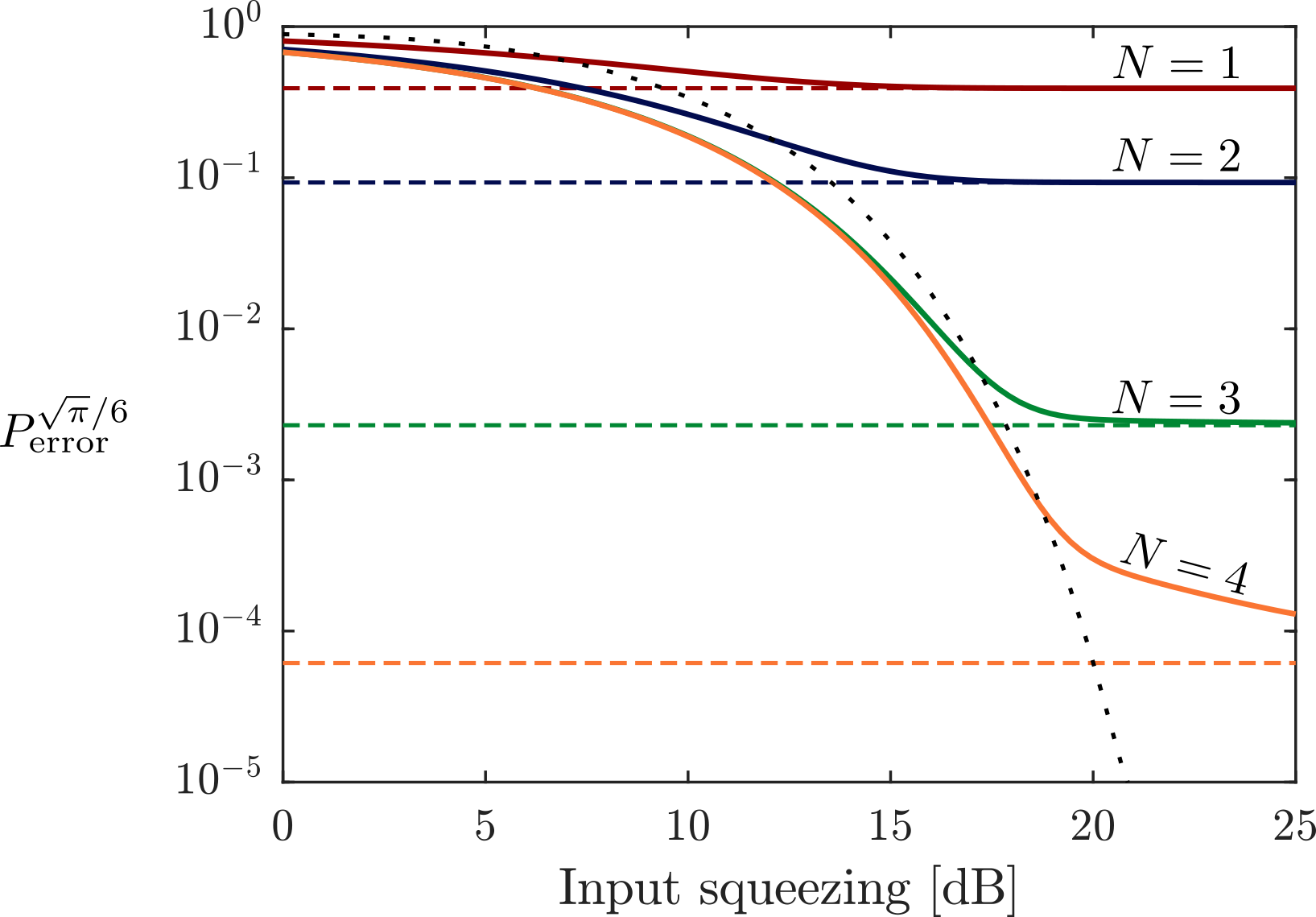}
    \caption{Effective shift error as a function of input squeezing. Dashed lines show the limits for infinite squeezing. The black dotted line shows the effective shift error for the approximate GKP states given by equation \eqref{eq:GKPapprox}.}
    \label{fig:Perror}
\end{figure}

As pointed out in \cite{tzitrin2019towards}, a low $P_\textrm{error}^{\sqrt{\pi}/6}$ is both sufficient and necessary for useful GKP states. Fig. \ref{fig:Perror} shows $P_\textrm{error}^{\sqrt{\pi}/6}$ of the states generated using our protocol. The tendency for smaller $P_\textrm{error}^{\sqrt{\pi}/6}$ for large $N$ and large input squeezing confirms that the generated states are indeed suitable for the GKP error correction protocol. For comparison, the dotted line shows $P_\textrm{error}^{\sqrt{\pi}/6}$ of the approximate grid states given by eq. \eqref{eq:GKPapprox}. For low input squeezing and high $N$ our states perform slightly better than the states of eq. \eqref{eq:GKPapprox}. This is because high $N$ states have low errors in the $P$ quadrature, independent of the input squeezing. However, since $P_\textrm{error}^{\sqrt{\pi}/6}$ depends on the quality of the state in both $X$ and $P$, we cannot keep improving it by solely increasing $N$, without also increasing the input squeezing, and vice versa, as seen from Fig. \ref{fig:Perror}.

\section{PREPARATION GATE INTERACTION STRENGTHS}\label{app:prep}
The preparation gates are used to shape the envelope of the prepared grid state in order to optimize the quality of the states. For infinite input squeezing, the coefficients of Eq. \eqref{eq:result} for $N\in{2,3,4}$ are:
\begin{itemize}
    \item $N=2$:
    \begin{align}
        c_1 &= \cos(\pi/4 + 2\sqrt{\pi}u_2) \nonumber\\
        c_2 &= \sin(\pi/4 + 2\sqrt{\pi}u_2)
    \end{align}
        \item $N=3$:
    \begin{align}
        c_1 &= \cos(\pi/4 + 4\sqrt{\pi}u_2)\cos(\pi/4 + 6\sqrt{\pi}u_3) \nonumber\\
        c_2 &= \cos(\pi/4 + 4\sqrt{\pi}u_2)\sin(\pi/4 + 6\sqrt{\pi}u_3) \nonumber\\
        c_3 &= \sin(\pi/4 + 4\sqrt{\pi}u_2)\cos(\pi/4 + 2\sqrt{\pi}u_3) \nonumber\\
        c_4 &= \sin(\pi/4 + 4\sqrt{\pi}u_2)\sin(\pi/4 + 2\sqrt{\pi}u_3)
    \end{align}
            \item $N=4$:
    \begin{align}
        c_1 =& \cos(\pi/4 + 8\sqrt{\pi}u_2)\cos(\pi/4 + 12\sqrt{\pi}u_3) \nonumber\\
        &\times \cos(\pi/4 + 14\sqrt{\pi}u_4) \nonumber\\
        c_2 =& \cos(\pi/4 + 8\sqrt{\pi}u_2)\cos(\pi/4 + 12\sqrt{\pi}u_3) \nonumber\\
        &\times \sin(\pi/4 + 14\sqrt{\pi}u_4) \nonumber\\
        c_3 =& \cos(\pi/4 + 8\sqrt{\pi}u_2)\sin(\pi/4 + 12\sqrt{\pi}u_3) \nonumber\\
        &\times \cos(\pi/4 + 10\sqrt{\pi}u_4) \nonumber\\
        c_4 =& \cos(\pi/4 + 8\sqrt{\pi}u_2)\sin(\pi/4 + 12\sqrt{\pi}u_3) \nonumber\\
        &\times \sin(\pi/4 + 10\sqrt{\pi}u_4) \nonumber\\
        c_5 =& \sin(\pi/4 + 8\sqrt{\pi}u_2)\cos(\pi/4 + 4\sqrt{\pi}u_3) \nonumber\\
        &\times \cos(\pi/4 + 6\sqrt{\pi}u_4) \nonumber\\
        c_6 =& \sin(\pi/4 + 8\sqrt{\pi}u_2)\cos(\pi/4 + 4\sqrt{\pi}u_3) \nonumber\\
        &\times \sin(\pi/4 + 6\sqrt{\pi}u_4) \nonumber\\
        c_7 =& \sin(\pi/4 + 8\sqrt{\pi}u_2)\sin(\pi/4 + 4\sqrt{\pi}u_3) \nonumber\\
        &\times \cos(\pi/4 + 2\sqrt{\pi}u_4) \nonumber\\
        c_8 =& \sin(\pi/4 + 8\sqrt{\pi}u_2)\sin(\pi/4 + 4\sqrt{\pi}u_3) \nonumber\\
        &\times \sin(\pi/4 + 2\sqrt{\pi}u_4) \nonumber\\
    \end{align}
    
\end{itemize}
and $c_m=c_{2^N-m+1}$ for $m>2^{N-1}$. Using these expressions one can tune the interaction strengths, $u_k$, to optimize the quality of the prepared grid state with respect to any desired figure of merit (FOM). In general, for the finite sum of $2^N$ infinitely squeezed states as given by Eq. \eqref{eq:result}, the expectation value of $\hat{D}(\sqrt{2\pi})$ used to calculate $\Delta_P$ (Eq. \eqref{eq:Delta}) is given by
\begin{equation}
    \langle\hat{D}(\sqrt{2\pi})\rangle = \frac{1}{2}\sum_{s=1}^{2^N-1}c^*_s c_{s+1},
\end{equation}
and the grid-state-basis wavefunction $\rho_{uv,uv}$ used to calculate $P_\textrm{error}^{\sqrt{\pi}/6}$ (Eq. \eqref{eq:Perror}) is given by
\begin{equation}
    \rho_{uv,uv}=2\pi^{-1/2}\Big\lvert\sum_{s=1}^{2^{N-1}} c_{s+2^{N-1}} \cos(2\sqrt{\pi}sv-\sqrt{\pi})\Big\rvert^2\delta(u),
\end{equation}
\begin{table*}[]
\caption{Optimal interaction strengths $u_k$, and resulting effective shift errors and effective squeezing for $N=1,2,3,4$. The optimal distribution is calculated by optimizing over all $c_k$ in Eq. \eqref{eq:result} and the flat distribution is given by equal $c_k$'s, which is obtained by removing the ''preparation gate" in our protocol.}
\label{table:I}
\begin{tabular}{ll||l|l|l||l|l|l||l}

                          &            & \multicolumn{3}{c||}{Optimize $P_\textrm{error}^{\sqrt{\pi}/6}$}                                                                                                                             & \multicolumn{3}{c||}{Optimize $\Delta_P$}                                                                                                                                               &                                 \\ \hline
                          \hline
\multicolumn{1}{l|}{$N$} & FOM                                                                                     & $[u_1,..,u_N]$            & This work                                                                & \multicolumn{1}{c||}{\begin{tabular}[c]{@{}c@{}}Optimal \\ distribution\end{tabular}} & $[u_1,..,u_N]$          & This work                                                              & \multicolumn{1}{c||}{\begin{tabular}[c]{@{}c@{}}Optimal\\ distribution\end{tabular}} & \multicolumn{1}{l}{Flat distribution}                                                      \\ \hline
\multicolumn{1}{l|}{1}   & \begin{tabular}[c]{@{}l@{}}$P_\mathrm{error}^{\sqrt{\pi}/6}$ \\ $\Delta_P$\end{tabular} & $[0]$                     & \begin{tabular}[c]{@{}l@{}}$3.9\times 10^{-1}$ \\ $6.6$ dB\end{tabular}  & \begin{tabular}[c]{@{}l@{}}$3.9\times 10^{-1}$\\ $6.6$ dB\end{tabular}               & $[0]$                   & \begin{tabular}[c]{@{}l@{}}$3.9\times10^{-1}$\\ $6.6$ dB\end{tabular}  & \begin{tabular}[c]{@{}l@{}}$3.9\times10^{-1}$\\ $6.6$ dB\end{tabular}               & \multicolumn{1}{l}{\begin{tabular}[c]{@{}l@{}}$3.9\times 10^{-1}$\\ $6.6$ dB\end{tabular}} \\ \hline
\multicolumn{1}{l|}{2}   & \begin{tabular}[c]{@{}l@{}}$P_\mathrm{error}^{\sqrt{\pi}/6}$ \\ $\Delta_P$\end{tabular} & $[0,0.045]$               & \begin{tabular}[c]{@{}l@{}}$9.3\times10^{-2}$\\ $11.6$ dB\end{tabular}   & \begin{tabular}[c]{@{}l@{}}$ 9.3\times10^{-2}$\\ $11.6$ dB\end{tabular}              & $[0,0.093]$             & \begin{tabular}[c]{@{}l@{}}$9.7\times10^{-2}$\\ $11.7$ dB\end{tabular} & \begin{tabular}[c]{@{}l@{}}$9.7 \times 10^{-2}$\\ $11.7$ dB\end{tabular}            & \multicolumn{1}{l}{\begin{tabular}[c]{@{}l@{}}$1.2\times10^{-1}$\\ $10.4$ dB\end{tabular}} \\ \hline
\multicolumn{1}{l|}{3}   & \begin{tabular}[c]{@{}l@{}}$P_\mathrm{error}^{\sqrt{\pi}/6}$ \\ $\Delta_P$\end{tabular} & $[0,0.053,0.033]$         & \begin{tabular}[c]{@{}l@{}}$2.3\times 10^{-3}$ \\ $16.6$ dB\end{tabular} & \begin{tabular}[c]{@{}l@{}}$2.1\times10^{-3}$\\ $16.6$ dB\end{tabular}               & $[0,0.040,0.026]$       & \begin{tabular}[c]{@{}l@{}}$6.7\times10^{-3}$\\ $17.0$ dB\end{tabular} & \begin{tabular}[c]{@{}l@{}}$7.6\times 10^{-3}$\\ $17.0$ dB\end{tabular}             & \multicolumn{1}{l}{\begin{tabular}[c]{@{}l@{}}$7.8\times10^{-2}$\\ $13.7$ dB\end{tabular}} \\ \hline
\multicolumn{1}{l|}{4}   & \begin{tabular}[c]{@{}l@{}}$P_\mathrm{error}^{\sqrt{\pi}/6}$ \\ $\Delta_P$\end{tabular} & $[0,0.038, 0.027, 0.015]$ & \begin{tabular}[c]{@{}l@{}}$6.1\times 10^{-5}$\\ $20.6$ dB\end{tabular}  & \begin{tabular}[c]{@{}l@{}}$5.1\times10^{-7}$\\ $19.9$ dB\end{tabular}               & $[0,0.024,0.015,0.008]$ & \begin{tabular}[c]{@{}l@{}}$1.8\times10^{-3}$\\ $22.3$ dB\end{tabular} & \begin{tabular}[c]{@{}l@{}}$1.3\times 10^{-3}$\\ $22.6$ dB\end{tabular}             & \multicolumn{1}{l}{\begin{tabular}[c]{@{}l@{}}$3.3\times10^{-2}$\\ $16.9$ dB\end{tabular}} \\ \hline
\hline
\end{tabular}
\end{table*}
\noindent assuming symmetry around $X=0$, i.e. $c_m=c_{2^N-m+1}$. 

Table \ref{table:I} shows the calculated optimal interaction strengths and obtained FOMs when optimizing $P_\textrm{error}^{\sqrt{\pi}/6}$ and $\Delta_P$ respectively. For comparison, we also compute the FOMs for the unconstrained optimal distribution, i.e. when freely optimizing over all possible coefficients $c_k$. We see, that even though we have only $N$ degrees of freedom to tune $2^N-1$ independent variables, we can obtain practically optimal FOMs. The only notable difference is for $N=4$ when optimizing $P_\textrm{error}^{\sqrt{\pi}/6}$, for which the optimal distribution is two orders of magnitude better than what can be obtained in our scheme. However, for practical purposes, the quality of the generated states will be limited by external noise sources, and therefore the optimal number of rounds will likely be limited to $N=3$ (see Appendix \ref{app:realisticnoise}). 

We also list the values of the FOMs corresponding to a flat distribution, i.e. where all $c_k$'s are equal. Such a distribution is obtained in our scheme by removing the preparation gates. For $N\geq3$ we see a significant improvement by tuning the peaks compared to the flat distribution. 

For comparison, the technique of adaptive phase estimation \cite{terhal2016encoding} also prepares a finite superposition of squeezed states. For $M=7$ rounds of phase estimation, $8$ peaks are obtained corresponding to $N=3$ in our scheme. The quality of the states obtained depends on measurement results, but the best-case scenario for $M=7$ has $P_\textrm{error}^{\sqrt{\pi}/6}=4.1\times10^{-3}$ and $\Delta_P=16.1$ compared to $P_\textrm{error}^{\sqrt{\pi}/6}=2.3\times10^{-3}$ and $\Delta_P=16.6$ for $N=3$ of our protocol. Our protocol thus outperforms adaptive phase estimation for any measurement outcome. However, the phase estimation protocol has the advantage of generating one peak at a time, and therefore the number of rounds can be better tuned to optimize the output state given the noise of the particular system. Yet, we believe that the main advantage of our protocol is the lack of measurements, which ultimately significantly speeds up the protocol, resulting in less accumulated noise. Additionally, for trapped-ion systems, qubit measurements can disturb the bosonic mode depending on the outcome, thus requiring postselection even with the phase estimation protocol. Therefore phase estimation does not scale well in this system, as the success probability decreases exponentially with the number of rounds. 

We generally note that $P_\textrm{error}^{\sqrt{\pi}/6}$ is a much more sensitive FOM compared to $\Delta_P$. For example, for $N=4$ the effective shift error can be improved by several orders of magnitude by tuning the coefficients, whereas the effective squeezing is ``only" improved by less than 5 dB, i.e. less than an order of magnitude. Note also, that for $N=3$ the difference in $\Delta_P$ when optimizing $\Delta_P$ compared to $P_\textrm{error}^{\sqrt{\pi}/6}$ is only 0.4 dB, where the difference in $P_\textrm{error}^{\sqrt{\pi}/6}$ is approximately a factor 3. This shows that states with similar effective squeezing can have significantly different effective shift errors. Since the effective shift error is more directly related to the error-correcting properties of the GKP states, care should therefore be taken when comparing only effective squeezing parameters.

For the results presented in the main text of this paper we have chosen $u_k$'s to optimize $P_\textrm{error}^{\sqrt{\pi}/6}$. For finite squeezing we have found no noticeable improvement in the FOMs by further tuning the $u_k$'s to take into account the finite squeezing, and we have therefore used the values of $u_k$ listed in Table \ref{table:I} for all simulations.  

\section{NO INITIAL SQUEEZING}\label{app:nosqueeze}
\begin{figure}
    \centering
    \includegraphics{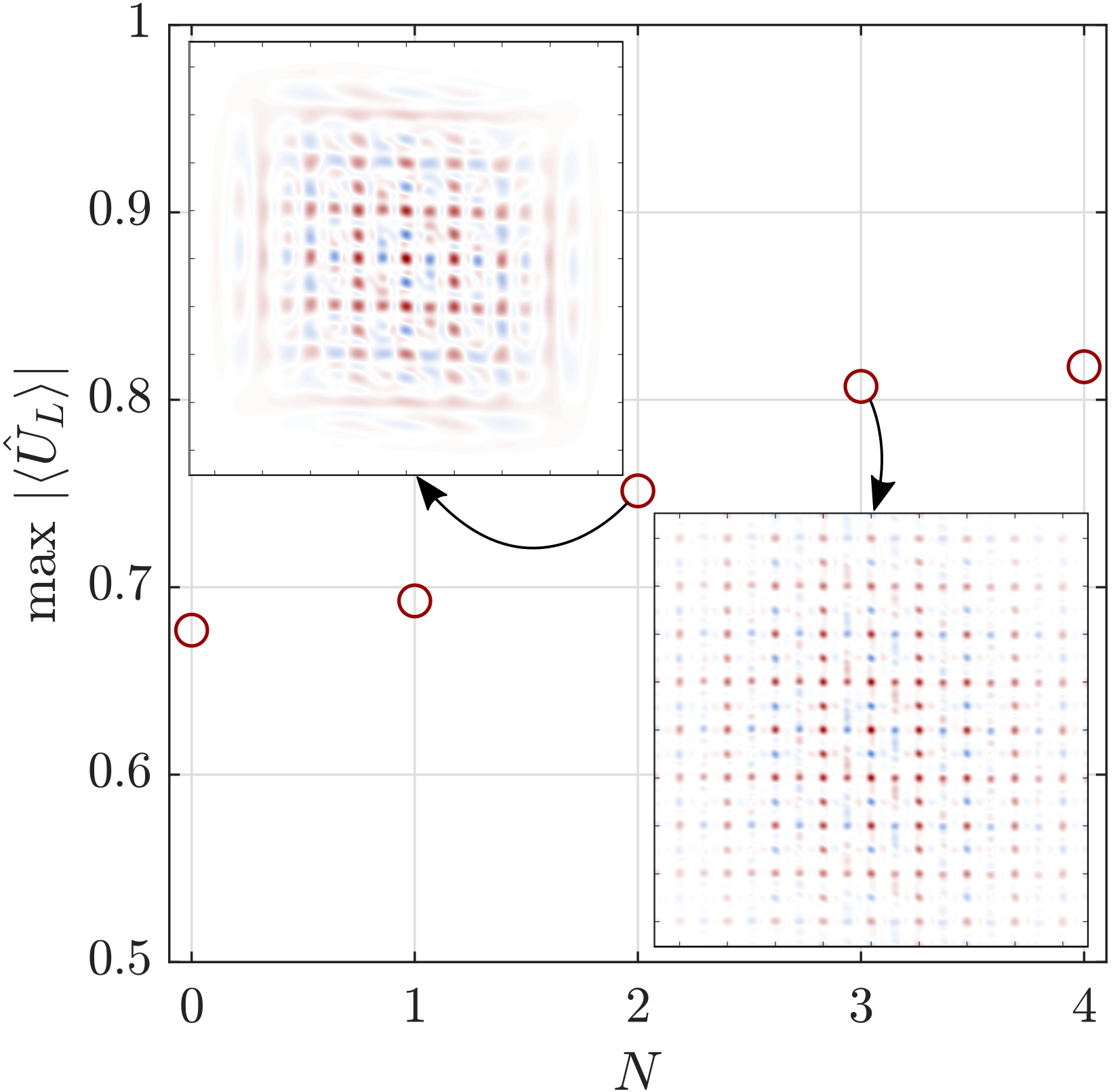}
    \caption{Maximum expectation value over all rotated Pauli-operators in the GKP-logical subspace for states generated without initial squeezing. The states are generated by applying the protocol of Fig. \ref{fig:scheme}(a) twice with a $\pi/2$ rotation of the bosonic mode in between. As $N$ increases the expectation value does not converge to 1, showing that these states do not represent pure GKP logical states. The insets show the Wigner functions of the generated state for $N=2$ and $N=3$.}
    \label{fig:NoSqueezing}
\end{figure}
In Fig. \ref{fig:result} we showed that high effective squeezing can be generated without any initial squeezing in the bosonic mode. This raises the question whether the protocol can be used to generate grid states directly from vacuum. In particular, one can start from vacuum and apply the protocol twice, once in each quadrature direction, to generate high effective squeezing in both quadratures simultaneously. The Wigner functions of the states generated with this approach are shown for $N=2$ and $N=3$ in the insets of Fig. \ref{fig:NoSqueezing}. The states have a clear grid-like structure, but it is not immediately clear if they represent useful GKP logic states, i.e. if they approach the form $c_0\ket{0}_\textrm{GKP} + c_1\ket{1}_\textrm{GKP}$ for some coefficients $c_0$ and $c_1$. To examine this, we calculate the expectation value of the rotated Pauli operators 
\begin{equation}
  \hat{U}_L=(|c_0|^2 - |c_1|^2)\hat{Z}_L + 2\Re(c_0'c_1)\hat{X}_L + 2\Im(c_0'c_1)\hat{Y}_L  .
\end{equation}
If the generated state approaches a logical GKP state we should have $|\langle\hat{U}_L\rangle|\rightarrow1$ for some $(c_0,c_1)$. Fig. \ref{fig:NoSqueezing} shows the maximum value of $|\langle\hat{U}_L\rangle|$ as a function of $N$. We observe that $|\langle\hat{U}_L\rangle|$ does not appear to converge to $1$, showing that the states do not represent useful GKP logic states. Still, The generated states do have high effective squeezing in both quadratures, so they would be a useful resource for detecting displacements \cite{duivenvoorden2017single}. 

\section{REALISTIC NOISE PARAMETERS}\label{app:realisticnoise}

\begin{figure}
    \centering
    \includegraphics{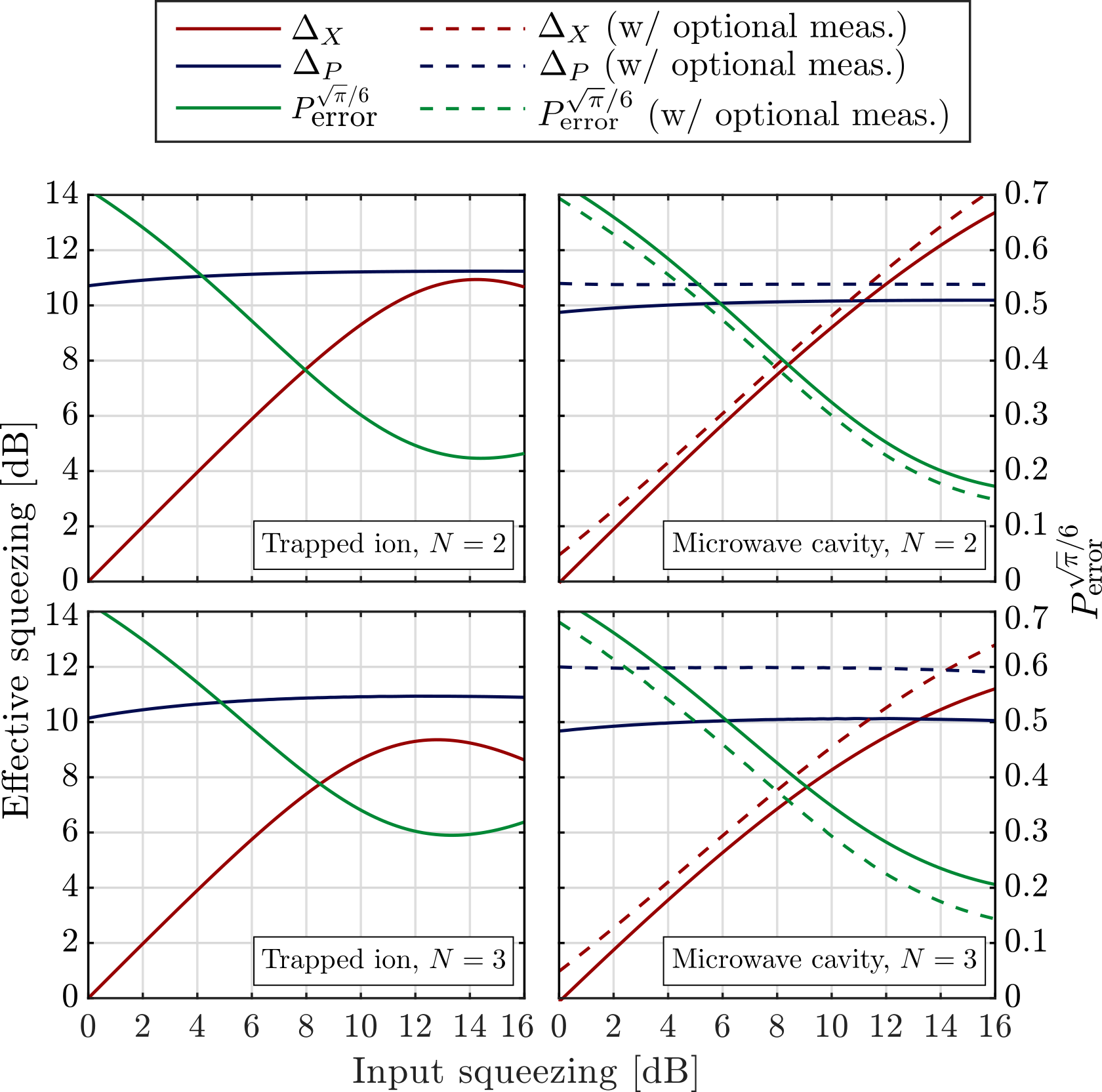}
    \caption{Effective squeezing and effective shift error as function of input squeezing as simulated using the master equation \eqref{eq:master}, using the noise types and strengths associated with recent experiments using trapped ions \cite{fluhmann2019encoding} and microwave cavities \cite{campagne2019stabilized}, as listed in Table \ref{table:II}.}
    \label{fig:realistic}
\end{figure}

Grid states were generated for the first time very recently in the motional state of a trapped ion \cite{fluhmann2019encoding} and in a microwave cavity field coupled to a superconducting circuit \cite{campagne2019stabilized}. These experiments obtained effective squeezing parameters of $\Delta_X=5.5$ dB and $\Delta_P=7.3$ dB, and $\Delta_X,\Delta_P\in[7.4;9.5]$ respectively. To benchmark our protocol, we simulate it with the Lindblad master equation, Eq. \ref{eq:master}, using the parameters given in these experiments, as listed in Table \ref{table:II}. The calculated FOMs as a function of input squeezing are shown in Fig. \ref{fig:realistic}. In both platforms we can obtain more than 10 dB effective squeezing in both quadratures simultaneously with $>11$ dB input squeezing. For the trapped ion platform  the optimal number of rounds is with $N=2$. For the microwave cavity platform both $N=2$ and $N=3$ allows for more than 10 dB output effective squeezing for sufficient input squeezing, with $N=3$ reaching 12 dB using the measurement strategy measurement as explained in the main text (dashed lines of Fig. \ref{fig:realistic}). 
\begin{table}[]
\caption{Interaction timescale and relevant noise types and rates in recent experiments with microwave cavities \cite{campagne2019stabilized} and trapped ions \cite{fluhmann2019encoding}. The interaction timescale, $T$, is the time required to perform the operations $e^{i\hat{X}\hat{\sigma}_y}$ or $e^{i\hat{P}\hat{\sigma}_x}$.}
\label{table:II}
\begin{tabular}{ll}

\multicolumn{2}{c}{Trapped ion \cite{fluhmann2019encoding}}                                                                                                               \\ \hline
\hline
\multicolumn{1}{c}{Interaction timescale} & \multicolumn{1}{l}{ $T=11$ $\mu$s}                                                                                        \\ \hline
\multicolumn{1}{c}{Phonon dephasing}           & \multicolumn{1}{l}{\begin{tabular}[c]{@{}l@{}}$\gamma^{-1} = 140$ ms\\ $\gamma T=7.7\times10^{-5}$\end{tabular}}     \\ \hline
\hline
                                                 &                                                                                                                       \\ 
\multicolumn{2}{c}{Microwave cavity \cite{campagne2019stabilized}}                                                                                                                                 \\ \hline
\hline
\multicolumn{1}{c}{Interaction timescale} & \multicolumn{1}{l}{$T=0.34$ $\mu$s}                                                                                      \\ \hline
\multicolumn{1}{c}{Qubit decay}                & \multicolumn{1}{l}{\begin{tabular}[c]{@{}l@{}}$\gamma^{-1} = 50$ $\mu$s\\ $\gamma T=6.8\times10^{-3}$\end{tabular}}  \\ \hline
\multicolumn{1}{c}{Qubit dephasing}            & \multicolumn{1}{l}{\begin{tabular}[c]{@{}l@{}}$\gamma^{-1} = 60$ $\mu$s\\ $\gamma T=5.7\times10^{-3}$\end{tabular}}  \\ \hline
\multicolumn{1}{c}{Photon decay}               & \multicolumn{1}{l}{\begin{tabular}[c]{@{}l@{}}$\gamma^{-1} = 245$ $\mu$s\\ $\gamma T=1.4\times10^{-3}$\end{tabular}} \\ \hline
\hline

\end{tabular}
\end{table}

\end{appendix}

\bibliography{References}

\end{document}